\documentclass[aps,prb,reprint,groupedaddress]{revtex4-1}

\usepackage{graphicx}
\usepackage{amsmath}

\begin{document}
  
\title{Constraints on nanomaterial structure from experiment and theory: Reconciling partial representations}

\author{Vladan Mlinar}
\email[]{vladan$\_$mlinar@brown.edu}
\affiliation{School of Engineering, Brown University, Providence, RI 02912, USA}

\date{\today}

\begin{abstract}
To facilitate the design and optimization of nanomaterials for a given application it is necessary to understand the relationship between structure and physical properties. For large nanomaterials, there is imprecise structural information so the full structure is only resolved at the level of partial representations. Here we show how to reconcile partial structural representations using constraints from structural characterization measurements and theory to maximally exploit the limited amount of data available from experiment. We determine a range of parameter space where predictive theory can be used to design and optimize the structure. Using an example of variation of chemical composition profile across the interface of two nanomaterials, we demonstrate how, given experimental and theoretical constraints, to find a region of structure-parameter space within which computationally explored partial representations of the full structure will have observable real-world counterparts. 
\end{abstract}


\maketitle

\section{Introduction} 
\label{sec:intro}

The prerequisite for the design and optimization of nanomaterials for a given application is an understanding of the relationship between structure and physical properties~\cite{Shultz,Carter,Franceschetti,Tarantola_nature}. However, with large nanomaterials, on scales of several thousand atoms and more, there are large uncertainties and imprecise structural information~\cite{Efros_09,Mlinar_2012,Lopez,Mlinar_graphs}. This is caused by both the lack of atomic resolution of structural characterization methods, and limitations of predictive theory methods, such as those previously successfully applied in, e.g., molecular spectroscopy~\cite{Barone}, or on a few-atom scale to predict new materials~\cite{Ortiz,Chelikowsky_2}.

\begin{figure}
\includegraphics[width=1.0\linewidth]{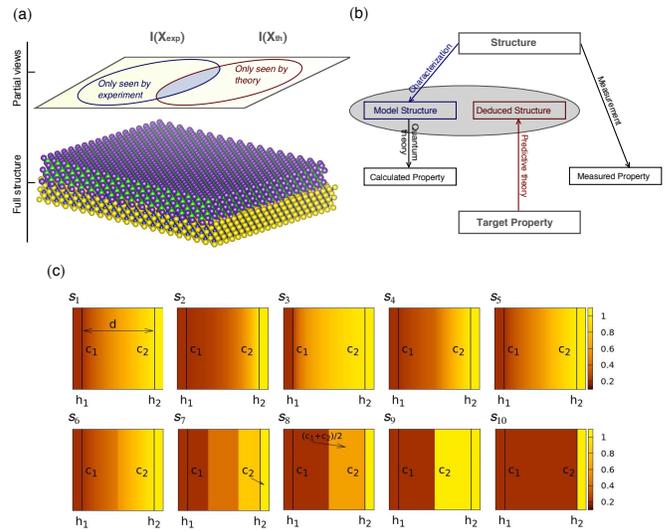}  
\caption{(a) The partial views of the structure, as ``seen'' by experiment and theory; (b) Description of the structure obtained from structural characterization, model structure, and deduced from predictive theory starting from a target physical property, deduced structure; (c) Different chemical composition profiles (CCP), $s_1, s_2, ..., s_{10}$, illustrating variation from linear to step-like. Arrow in $s_7$ points to region with concentration $c_2$ and in $s_8$, region with $(c_1+c_2)/2$.
\label{fig:fig.1}}                      
\end{figure}

For large nanomaterials, the full atomistic structure is represented only indirectly by descriptive quantities, ``motifs'' such as chemical composition profile (CCP), geometry, confining potential, etc. This leads to the loss of structural information. The full structural information, atoms and their positions, is replaced by the partial structural information contained in motifs. Whereas knowledge of the full information guarantees knowledge of partial representations of the structure via motifs, knowledge of partial representations is not, in general, sufficient to determine uniquely the underlying atomistic structure. Consequently, different partial representations of the same full structure can exist, e.g., as ``seen'' by experiment and theory, illustrated in Figure~\ref{fig:fig.1}~(a)~and~(b). Often, these representations are reconciled by ``cherry-picking'' one solution that fits all constraints, regardless of how loose they might be.

For example, it was shown that the interface in a core/shell CdZnSe/ZnSe nanocrystal was key in interpreting absence of intensity fluctuations in photoluminescence (PL) spectra, also known as blinking~\cite{Efros_09}. The electron micorgraph of a CdZnSe/ZnSe nanocrystal could not give precise information about the interface between CdZnSe core and ZnSe shell, so numerical calculations were used to find a possible composition profile at the interface that would eliminate blinking. 

In this case, structural information was deduced from two sources: (i) structural characterization measurements, which gave one partial structural representation, and (ii) predictive theory, which was based on the desired outcome (no blinking in the PL spectra). In order to get the best possible representation of the structure using employed motifs, these need to be reconciled. 

In this work, we show how to reconcile partial structural representations using constraints from structural characterization measurements and theory to maximally exploit the limited amount of data available from experiment. We determine a range of parameter space where predictive theory can be used to design and optimize the structure.

Using an example of variation of chemical composition profile across the interface of two nanomaterials, we demonstrate how, given experimental and theoretical constraints, to find a region of structure-parameter space within which computationally explored partial representations of the full structure will have observable real-world counterparts. 

This example is chosen because, in addition to enabling us to discuss the underlying ideas in an elegant way and without lost of generality (see below), the findings could be directly relevant to the analysis of e.g., diode lasers from epitaxial II-VI and III-V semiconductor heterostructures~\cite{Chai_APL,Mlinar_2013,Becla}. Interdiffusion at the heterointerface can lead to chemical composition changes of the interface, influencing optical properties, such as optical bandgap, and consequently device parameters~\cite{Chai_APL}. 

%
%

\section{Method}
\label{sec:meth}

  
\paragraph*{Model system.} 

Figure~\ref{fig:fig.1}~(c) shows all possible CCPs, denoted as $s_1$, $s_2$, ..., $s_{10}$. They represent variation of chemical composition across the interface, from a nanomaterial with composition $c_1$ to a nanomaterial with composition $c_2$. The length within which the variation of CCPs occurs is denoted as $d$ (see e.g., $s_1$ in Figure~\ref{fig:fig.1}~(c)). 

The variations of CCPs in Figure~\ref{fig:fig.1}~(c) are chosen such to reflect some of the most common profiles as extracted experimentally (e.g. those CCPs obtained from structural characterization measurements of ZnSe/ZnCdSe and assuming that diffusion follows Fick's law)~\cite{Chai_APL,Becla}, obtained by linking CCP with variation in PL shift~\cite{Chai_APL}, or assumed in theoretical model (e.g., abrupt variation of CCP across the interface)~\cite{Chai_APL,Mlinar_2012,Mlinar_2013}. 

Model CCPs are used to demonstrate the method and discuss implications of the findings. Links to real-world systems are mentioned in the next section where appropriate. 

\paragraph*{Basic logic.} 

Ideally, both, experiment and theory should be able to distinguish between each and every CCP. However, typically there are several CCPs that fit structural characterization measurements, and a range of CCPs acceptable by theory~\cite{Chai_APL,Mlinar_2012}. We reconcile CCPs as ``seen'' by theory and by experiment using Dempster-Shafer evidence theory~\cite{Dempster,Shafer,Parsons}.  The idea is to construct a so-called evidence structure using all the data we have from structural characterization measurements and theory, determine level of support and conflict between evidence from experiment and theory, and then apply a quantitative measure to determine which CCPs are acceptable by both experiment and theory.

\paragraph*{Construction of the evidence structure.}

We start from the set $S = \{s_1, s_2,..., s_{10}\}$ which contains all possible CCPs, as shown in Figure~\ref{fig:fig.1}~(c). There are $2^{10}-1$ possible (non-empty) subsets of $S$, e.g., subsets $\{s_1\}$, $\{s_1,s_2\}$, $\{s_1,s_2,s_3\}$, etc. Some subsets of CCPs from $S$ are allowed by structural characterization measurements data, $\{S_{P_E }\}$, and others by theoretical data, $\{S_{P_T }\}$. The same CCPs can belong to both $\{S_{P_E}\}$ and $\{S_{P_T }\}$ if they are identified by structural characterization measurements and theory. In the best case scenario, we should be able to distinguish between each and every CCP, and in the worst case, we would not be able to distinguish among any of CCPs from $S$, i.e., any $s_i$ from $S$ would be an acceptable solution. 

Any of the possible subsets contain some CCPs and have a certain probability of being valid. We characterize every subset $S_{P_i}$, where ($i=1,2,...,2^{10}$), by the mass of belief committed to it, $m(.)$, also called basic belief assignment (bba). Indeed, $m(.)$ represents our degree of belief in $S_{P_E}$ ($S_{P_T}$) given the evidence from experiment (theory). Mathematically, $m(.)$ is defined as $m(.): 2^{|S|}\rightarrow [0,1]$, i.e., mapping from the set of all subsets of $S$ (the power set of $S$), to [0,1] that satisfies the conditions: $m(\emptyset) = 0$ and $\sum_{X\in 2^{S}}m(X) = 1$ ~\cite{Dempster,Shafer}. 

Values of $m(.)$  are obtained based on the available evidence, and construction and application of the preference structure of the evidence as proposed in Ref. ~\cite{Wong}.  We generate quantitative $m_E$ based on the evidence from experimental results: (e1) the length over which there is variation of composition is $d$; (e2) the value of composition at $h_1$ is $c_1$; (e3) the value of composition at $h_2$ is $c_2$; (e4) $c_1 < c_2$; (e5) there is a gradient in CCP from $h_1$ to $h_2$; and (e6) there are no data to specify the exact form of the profile. 

Just as in the case of generating $m_E$, values of $m_T$ are obtained based on the evidence from predictive theory and by applying the preference structure of Ref. ~\cite{Wong}. From predictive theory, the structure is reconstructed given a physical/optical property [e.g., the peak PL energy, see also Figure~\ref{fig:fig.1}~(b)]. Given that, we can designate: (t1) value of composition at $(h_2-h_1)/2$ should be greater or equal to $(c_2-c_1)/2$; (t2) the value of composition at $h_1$ should be $< (c_2-c_1)/2$; (t3) the value of composition at $h_2$ should be $c_2$, and (t4) there is no constraint on whether there is an abrupt or gradient variation of the profile. 

Next, for each subset $S_P$, we introduce (and calculate) measures to quantify the strength of the evidence from both, experiment and theory, and the potential specific support that could be given to each $S_P$.  The belief, $Bel(S_P)$, determines the amount of support given to $S_P$, and is calculated as $Bel(S_P) = \sum_{X|X\subset S_P}m(X)$~\cite{Dempster,Shafer}. It actually measures the strength of the evidence; $Bel(S_P) = 0$ corresponds to the case of absence of evidence and $Bel(S_P) = 1$ denotes certainty. The plausibility, $Pl(S_P)$, determines the total amount of potential specific support that could be given to $S_P$, and is calculated as $Pl(S_P) = \sum_{X|S_P\cap X \neq\emptyset}m(X)$ ~\cite{Dempster,Shafer}. It actually represents the whole mass of belief that comes from all subsets of $S$ intersecting $S_P$. It also ranges from 0 to 1. Note that $Bel(S_P)\leq Pl(S_P)$~\cite{Shafer}.

\paragraph*{Reconciling partial representations.}

Finally, we apply the rule of combination to reconcile the experimental and theoretical partial representations~\cite{Dempster}. The Dempster-Shafer (DS) rule is expressed as 
$m_{DS}(S_P^{(DST)}) = [m_E\oplus m_T](S_P^{(DST)}) = m_{(E,T)}(S_P^{(DST)})/(1-K_{(E,T)})$, where $m_{(E,T)}(S_P^{(DST)}) = \sum_{S_{P1} \cap S_{P2} = S_P^{(DST)}}m_E(S_{P1})m_T(S_{P2})$
represents the conjunctive consensus on $S_P^{(DST)}$ between the experiment and theory, and $K_{(E,T)}= \sum_{S_{P1} \cap S_{P2} = \emptyset}m_E(S_{P1})m_T(S_{P2})$ is the total degree of conflict between the experiment and theory~\cite{Dempster,Shafer,Tessem,Parsons}. By calculating and properly interpreting the measures of the evidence theory, $Bel(.)$, $Pl(.)$, $m_{DS}(.)$, we are able to (i) reconcile CCPs that fit both available theoretical and experimental data, and (ii) identify region in structure parameter space for computer design and optimization of nanomaterials. 

\section{Results and Discussion}
\label{sec:res}

\begin{figure}  
\includegraphics[width=1.00\linewidth]{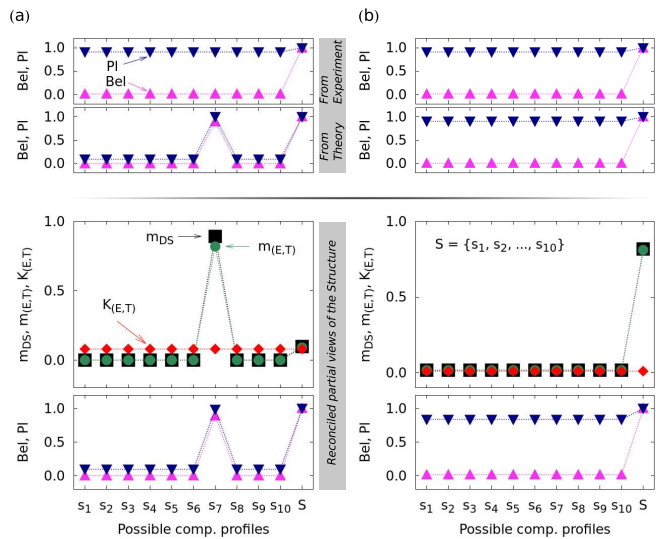}  
\caption{(a) The calculated amount of support, $Bel(.)$, and the calculated total amount of potential specific support, $Pl(.)$, that could be given to a single $s_i$ and the full set $S$. According to experiment, any CCP from the set $S$ is acceptable, but theory singles-out one CCP, $s_7$. Reconciling the two partial representations of the structure, where the Dempster-Shafer rule of combination is used to calculate combined bba $m_{DS}$,  the conjunctive consensus on $S_P^{(DST)}$, $m_{(E,T)}$, and the total degree of conflict $K_{(E,T)}$. (b) The same as (a), but for the case where any CCP from $S$ is acceptable both according to the structural characterization measurements and theory.
\label{fig:fig.2}}      
\end{figure}

We use CCPs from Figure~\ref{fig:fig.1}~(c) as input to our method. We consider three scenarios depending on the strength of the evidence on CCPs from experiment and theory. In the first scenario, we start from the premise that structural characterization measurements cannot distinguish between CCPs from Figure~\ref{fig:fig.1}~(c). Our findings are shown in Figure~\ref{fig:fig.2}. In the second and third scenarios, we (i) show how partial representations are reconciled with additional constraints introduced; and (ii) demonstrate how experimental and theoretical CCPs can be judged based on the evidence theory measures, the conjunctive consensus, $m_{(E,T)}$, and the total degree of conflict between experiment and theory, $K_{(E,T)}$.   

Figure~\ref{fig:fig.2}~(a) top shows calculated $Bel(.)$ and $Pl(.)$ for our first case: According to the structural characterization measurements, any CCP from the set $S$ is acceptable, which gives $Bel(s_i) = 0$, where $i = 1, 2,..., 10$, as there is no experimental evidence to support/single out any specific $s_i$. $Bel(S)=Pl(S) =1$, as, by definition, resulting CCP belongs to $S$. The theoretical prediction singles out one CCP, $s_7$, which gives $Bel(s_7 )\approx Pl(s_7 )=1$. Also, $Bel(S) =Pl(S) = 1$ because resulting CCPs always belong to $S$. The partial structural representations from the two sources, structural characterization and theory, are reconciled by using DS-rule to calculate $m_{DS}$, Figure~\ref{fig:fig.2}(a) bottom; $m_{DS}$ has maximum value for $s_7$ because the lack of constraints from experiment allows for any theoretical prediction to be accepted. The same conclusion can be also obtained by looking at consensus and conflict between experimental and theoretical data for different CCPs; $m(E,T) \sim m_{DS}$, and $K(E,T)\sim 0$. Calculated $Pl(.)$, after DS rule was applied, shows (Figure~\ref{fig:fig.2}(a) bottom) that $Pl(s_7) =Pl(S) = 1$, and approximately zero otherwise. Furthermore, calculated strength of evidence shows that resulting CCP certainly belongs to $S$, i.e., $Bel(S) = 1$, but it is likely $s_7$. Indeed, this is how the structural analysis of the blinking in the PL spectra of core/shell CdZnSe/ZnSe nanocrystal from Ref.~\cite{Efros_09}, mentioned above, should have been regarded.

Consequently, we can interpret these findings from a different perspective, if we could resolve the structure on the atomistic level using theory and provide input to experiment, e.g., to guide fabrication/synthesis, obviously, we would still lack the feedback from the experiment of how the real-world structure actually looks like. 

Figure~\ref{fig:fig.2}(b) shows the results for the case where both experimental and theoretical predictions are unconstrained. In that case, as intuitively clear, it can only be concluded that any CCP from $S$ could be a solution, but nothing more specific.

\begin{figure} 
\begin{center}
\includegraphics[width=0.8\linewidth]{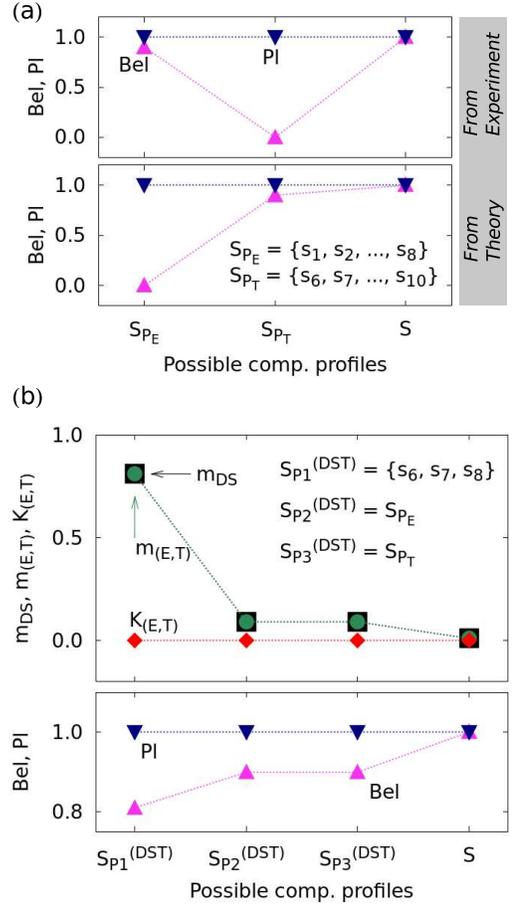}  
\caption{(a) The calculated amount of support, $Bel(.)$, and the calculated total amount of potential specific support, $Pl(.)$, that could be given to a $S_{P_E}$ and $S_{P_T}$. According to experiment, any CCP in $S_{P_E}$ is acceptable, and according to theory, any CCP in $S_{P_T}$; (b) Reconciling the two partial representations of the structure, $S_{P_E}$, $S_{P_T}$, with the corresponding $m_E(.)$ and $m_T(.)$. The Dempster-Shafer (DS) rule of combination is used to calculate $m_{DS}$, which identifies the subset of CCPs, $S_{P1}^{(DST)}$, that is acceptable by both, structural characterization and theory. 
\label{fig:fig.3}}      
\end{center}
\end{figure}

Next, we impose additional constraints, i.e., precise data from structural characterization measurements and theoretical predictions. The calculated $Bel(.)$ and $Pl(.)$ are shown in Figure~\ref{fig:fig.3}(a). To obtain the results within our model, and understand how they should be interpreted we construct the following experimental and theoretical premises. 

\emph{(i) Structural characterization measurements} can detect variation of CCP across the interface (where diffusion has been often assumed to follow Fick's law), but not resolve the gradient~\cite{Chai_APL,Mlinar_2012}. For our test study, starting from the initial set $S$, this gives a subset of CCPs, $S_{P_E}={s_1,s_2,...,s_8}$. We then obtain $Bel(S_{P_E})^{(Exp.)}\approx Bel(S)^{(Exp.)}=1$ and $Bel(S_{P_T})^{(Exp.)}=0$. Given that $Pl(.)$ represents support coming from all subsets of $S$ intersecting with a given subset, $Pl(S_{P_E})^{(Exp.)}=Pl(S_{P_T})^{(Exp.)}=Pl(S)^{(Exp.)}=1$. For example, for the subset $S_{P_E}$, $Pl(S_{P_E})$ counts support from $S_{P_E}$, $S_{P_T}$, and $S$. 

\emph{(ii) As allowed by theory}, different CCPs can be obtained by linking the variation of CCP across the interface with the shift of the peak PL energy for the sample of interest~\cite{Chai_APL}. Thus, for our case study, from the initial set $S$, we select CCPs given in $S_{P_T} =\{s_6,s_7,...,s_10\}$.  This gives $Bel(S_{P_T})^{(Th.)}\approx Bel(S)^{(Th.)}=1$, $Bel(S_{P_E})^{(Th.)}=0$, and $Pl(S_{P_E})^{(Th.)}=Pl(S_{P_T})^{(Th.)}=Pl(S)^{(Th.)}=1$.  

Figure~\ref{fig:fig.3} (b) shows $m_{DS}$ calculated using the DS rule of combination. The maximum value of $m_{DS}$ is used to identify the targeted subset of CCPs, which is found to be $S_{P1}^{(DST)}$. This means that in the parameter space of the partial structural representations, i.e., CCPs and the corresponding values they may take, there is a subset of CCPs ($S_{P1}^{(DST)}$) that is accepted by both experiment and theory. Also, after DS rule is applied, $Bel(S_{P1}^{(DST)}) < Bel(S_{P2}^{(DST)}) = Bel(S_{P3}^{(DST)}) < Bel(S)$. This means that it is more likely to find a targeted CCP in $S_{P2}^{(DST)}$ and/or $S_{P3}^{(DST)}$ than in $S_{P1}^{(DST)}$, as $S_{P1}^{(DST)} \subset S_P2^{(DST)}$ and $S_{P1}^{(DST)}\subset S_{P3}^{(DST)}$. Identified subset of CCPs, $S_{P1}^{(DST)}$, with three CCPs in Figure~\ref{fig:fig.3}~(b), represents a tentative solution based on the currently available experimental and theoretical evidence. With further evidence introduced, e.g., additional measurements, the set of acceptable CCPs can change. If we seek the true solution, then we should have full structural information; i.e., atoms and their positions. 

From a broader perspective, $S_{P1}^{(DST)}$ actually gives us a region of the structural parameter space within which explored and/or computationally designed partial structural representations of the structure can have observable real-world counterparts. Thus, by computationally exploring/searching the parameters within the space identified by the DS rule, we are able to provide input to fabrication/synthesis, useful from a design and optimization viewpoint, because it can be registered by the structural characterization and provide relevant feedback to the theory. 

\begin{figure}
\includegraphics[width=0.9\linewidth]{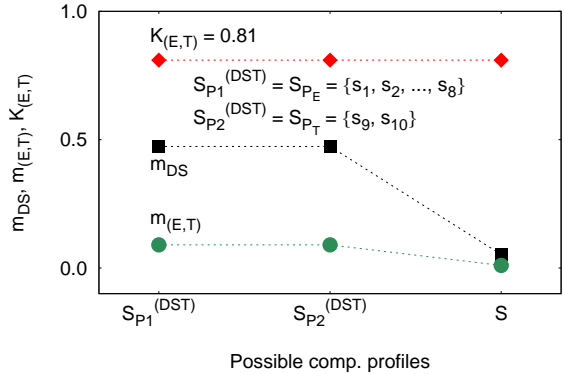}  
\caption{Reconciling partial representations in the case of variation of CCPs across the interface where, according to experimental results, $S_{P_E} = \{s_1, s_2,...,s_8\}$, but abrupt variation of CCP between $h_1$ and $h_2$, $S_{P_T} = \{s_9, s_{10}\}$; this corresponds to the introduction of geometric motifs. The calculated total degree of conflict, $K_{(E,T)} = 81\%$.
\label{fig:fig.4}}      
\end{figure}

Furthermore, arbitrarily selecting some CCPs (out of all that are allowed by calculations) can have severe consequences. If we retain $S_{P_E}$ from Figure~\ref{fig:fig.3}, but theoretically considered only those CCPs with an abrupt variation between the two nanomaterials, $S_{P_T}^{(Fig.~\ref{fig:fig.4})}=\{s_9,s_{10} \} \subset S_{P_T}^{(Fig.~\ref{fig:fig.3})}$, then the implications of (arbitrarily) narrowing down $S_{P_T}$ can be traced by calculating the total degree of conflict between the experimental and theoretical evidence, $K_{(E,T)}$. Our findings are shown in Figure~\ref{fig:fig.4}. Unlike the cases considered in Figure~\ref{fig:fig.2} and Figure~\ref{fig:fig.3}, where $K_{(E,T)}\sim 0$, we see that for the case in Figure~\ref{fig:fig.4}, $K_{(E,T)}$ is increased dramatically, to 81$\%$. 

An interesting and simple example of an arbitrary narrowing of the parameter space can be found when geometric motifs, such as the width of a quantum well, are introduced. One simply separates one material (barrier) from the other (well), basically introducing a sharp variation in CCP [see $s_9$ and $s_{10}$ in Figure~\ref{fig:fig.1}(c) and Figure~\ref{fig:fig.4}]. Clearly, a CCP fully describes the structure at this level of representation, so usage of additional geometrical motifs is not necessary. However, the implications of introducing geometric motifs can be significant. 

This approach and analysis are not limited to II-VI heterointerfaces, but may include e.g., materials with interfaces between different phases, such as solid-liquid or liquid-gas interfaces~\cite{Fadley}. Actually, the approach is valid for any nanomaterial-based system for which we do not know (and cannot extract) the full structure, but can only access partial structural representations via structural characterization and calculations. 

By reconciling partial structural representations using \emph{available} constraints from experiment/measurements and theory/calculations, we provide \emph{the upper limit} of what can be concluded about the structure given these inputs. Compared to full structural information, atoms and their positions, the upper limit constructed from partial representations is bounded by uncertainties inherent for structural characterization measurements, and numerical model. 

There are both experimental and theoretical benefits of this approach. \emph{Theoretically}, the initial parameter space of possible structures can be too large to be tractable, and arbitrarily choosing a region of parameter space might be misleading. By reconciling different possible CCPs, i.e., partial representations, such as those seen by experiment and theory, we maximally exploit the limited amount of data available from experiment, and determine a range of parameter space where predictive theory can be used to design and optimize the structure. \emph{Experimentally}, reconciling partial structural representations using experimental and theoretical  is important because it is very difficult (and certainly not recommended) to construct a large database from experimental results (which would require fabrication, characterization, then measurement of properties). Even if that were possible, the problem of imprecise structural information could not be eliminated. 

\section{Summary}
\label{sec:summ}

To summarize, we showed how to reconcile partial representations of the structure using constraints from experiment and theory and determined a range of parameter space where predictive theory can be used to design and optimize the structure. We provided the upper limit of what can be concluded about the structure given the inputs from structural characterization and theory. 

\section*{Acknowledgements}

This work was supported by NASA EPSCoR and a 2013 Brown University Seed Grant.


\end{document}